# Phase transition and enhanced hardness of LaB$_4$ under pressure

Chen Pan,[1] Sheng-Yan Wang,[1] Hui Tang,[1] Hong-Yun Wu,[1] Hong Jiang,[1] Yue-Hua Su,[1] and Chao Zhang[1*]

[1] *Department of Physics, School of Opto-electronic Information Science and Technology, Yantai University, Yantai, 264005, China*

The crystalline structures of lanthanum tetraboride, LaB$_4$, under pressure are investigated using a genetic algorithm method coupled with first-principles calculations. We find two low-enthalpy phases for LaB$_4$ as the thermodynamic ground state up to 200 GPa: a phase previously observed in experiments (*P*4/*mbm*) and a novel orthorhombic phase (*Pnma*-I). The *P*4/*mbm* phase transforms to the *Pnma*-I phase at around 106 GPa, accompanied by breakage of the B-octahedron. The *P*4/*mbm* phase of LaB$_4$ exhibits a high hardness (30.5 GPa) at ambient conditions, which is comparable to that of the hard material B$_4$C. The hardness of the two low-enthalpy phases increases with pressure. The electronic band structures shows that all of the competitive phases (two stable and two metastable phases) are metallic. Phonon calculations shows the *P*4/*mbm* and *Pnma*-I phases are thermodynamically stable in their respective pressure ranges. This elucidation of the phase transition behavior and hardness of LaB$_4$ provides new knowledge on the interesting high-pressure behaviors of the rare-earth metal borides.

*Corresponding author

Email address: phyczhang@ytu.edu.cn

1. Introduction

The lanthanum-boron system has attracted increasing interest as it holds potential for a variety of technologies applications.[1-4] There are two stable compounds in the lanthanum-boron system: lanthanum hexaboride ($LaB_6$) and lanthanum tetraboride ($LaB_4$).[5] Both are boron-rich solids, in which the boron packing crucially determines their distinct electronic, optical, and mechanical properties. $LaB_6$ has been extensively studied as an excellent thermionic electron emitter due to its low work function, long service life, high melting point, high emission density, high electrical conductivity, and high brightness.[6-11] $LaB_6$ is a monovalent metal characterized by hardness and a refractory nature, and serves as a standard reference material for X-ray powder diffraction (XRD).[12, 13] Unfortunately, compared to $LaB_6$, $LaB_4$ has only been sparsely visited by the scientific community.[14-20] The crystal structure of $LaB_4$ was resolved by XRD at ambient conditions, and is isostructural with $UB_4$, $CeB_4$, and $ThB_4$.[15, 17] The thermodynamic properties of $LaB_4$, such as its thermal expansion coefficient,[15] Debye temperature,[18] enthalpy,[18] and low-temperature heat capacity,[20] have also been investigated.

Pressure is a clean and efficient approach to tune the chemical and physical properties of a range of materials. Most materials undergo phase transitions and transform into new phases with novel properties when pressure is applied. Teredesai *et al*. investigated the high-pressure phase transitions of $LaB_6$ using Raman spectroscopy and X-ray diffraction in the pressure range of 0 – 20 GPa.[21] They found that $LaB_6$ transforms from a cubic phase to an orthorhombic phase at around 10 GPa. *Ab initio* electronic calculations indicated this structural phase transition is induced by an electronic band minimum intercepting the Fermi level. However, in subsequent Raman spectroscopy and X-ray diffraction experiments at room temperature, Dodwal *et al*. did not observe any phase transition up to 25 GPa.[22] Moreover, the previously reported subtle electronic phase transition that occurs around 10 GPa was not observed. Chao *et al*. used first-principles calculations to investigate the effect of pressure on the electronic structure, phonon spectrum, and optical properties of $LaB_6$.[23] The phonon spectra suggested the absence of phonon softening phenomenon in the 0 – 10 GPa pressure range, indicating there is no structural phase transition around 10 GPa. A study of $LaB_6$ showed its optical properties change suddenly at 45 GPa, and $LaB_6$ exhibits good thermal insulation performance from 0 – 35 GPa. The structural stability of $LaB_6$ under pressure was further confirmed by Li *et al*. using X-ray diffraction.[24] Moreover, the work functions of the (110) and (100) surfaces of $LaB_6$ were determined in the 0 – 39.1 GPa pressure range.

However, our knowledge of $LaB_4$—especially the effects of pressure on its properties—is relatively scarce compared to that of $LaB_6$. In this study, we investigated the behavior of $LaB_4$ up to 200 GPa using first-principles calculations based on the density functional theory. $LaB_4$ is predicted to transform from a tetragonal structure with *P*4/*mbm* symmetry to an orthorhombic structure with *Pnma* symmetry around 106 GPa. The *P*4/*mbm* phase of $LaB_4$ possesses excellent mechanical properties. The hardness of $LaB_4$ is comparable to that of $B_4C$, which is a hard material. The hardness of the *P*4/*mbm* phase increases with pressure and exhibits a large jump at the phase boundary.

2. Computational methods

Crystalline structure searches were conducted for $LaB_4$ by global optimization of free energy surfaces via the genetic algorithm (GA) technique implemented in the adaptive genetic algorithm (AGA) code.[25-29] We searched for stable structures at ambient pressure and high pressure using 1 – 4 formula units (fu) for each simulation cell at 0, 50, 100, 150, and 200 GPa. In the GA searches, we used the Vienna Ab initio Simulation Package (VASP) to relax the structures of the offspring structures in every GA generation by first-principles calculations within the density functional theory (DFT) framework;[30] we adopted the Perdew-Burk-Ernzerhof (PBE)[31] functional at the level of the generalized gradient approximation (GGA). The projected-augmented wave (PAW) method [32] was used with $5s^26p^65d^16s^2$ and $2s^22p^1$ as the valence electrons for La and B atoms, respectively. The electronic wave function was expanded using a plane wave basis set (kinetic energy cut-off, 520 eV) and Brillouin zone integrations were conducted using a Monkhorst-Pack [33] *k* mesh with $2\pi \times 0.02$ Å$^{-1}$ spacing. We optimized all candidate structures until the net forces on each atom were less than 1 meV/Å. A supercell approach and small displacement method via PHONONY code were employed to perform phonon calculations.[34] VASP was used to calculate the real-space force constants for a $1 \times 1 \times 2$ supercell of the *P*4/*mbm* phase and $2 \times 2 \times 1$ supercells of the other three phases of $LaB_4$.

## 3. Results and discussion

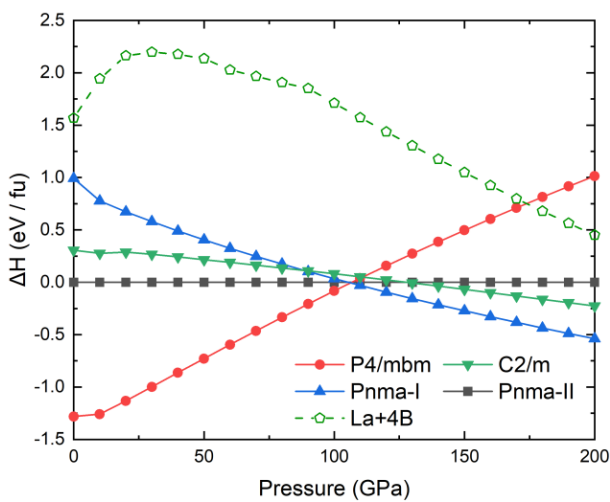

**Figure 1.** Enthalpy difference versus pressure for competitive structures of LaB$_4$, referenced to the *Pnma*-II phase.

A genetic algorithm method was employed to systematically investigate the phase stability of LaB$_4$ over the 0 – 200 GPa pressure range. The experimentally observed ambient-pressure tetragonal *P*4/*mbm* structure was correctly reproduced using only knowledge of the chemical composition, which confirmed the validity of using the GA technique to perform structural searches for LaB$_4$. In addition to the previously confirmed *P*4/*mbm* structure, three competitive structures were selected from the large variety of structures generated by the structural searches. The enthalpy difference of these competitive phases relative to a reference phase over the range from 0 – 200 GPa are shown in Figure 1, in which we indicate a possible decomposition (La + 4B) with dashed lines.[35-41]

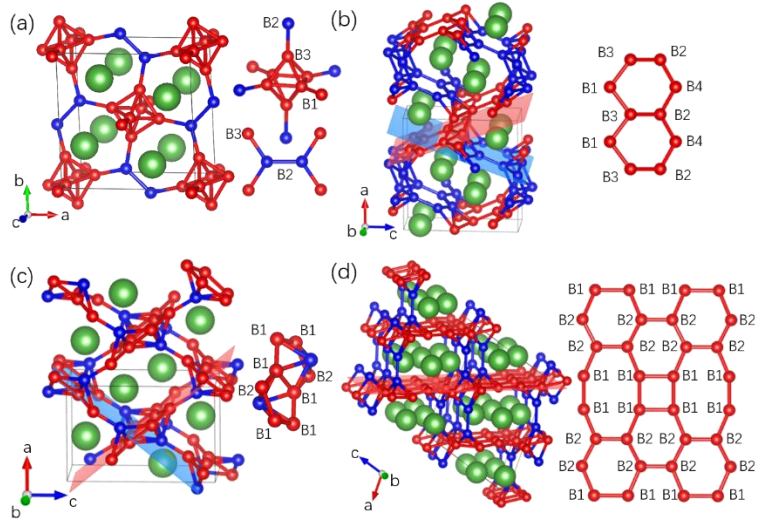

**Figure 2.** Crystal structures of LaB$_4$ for (a) *P*4/*mbm*, (b) *Pnma*-I, (c) *Pnma*-II, and (d) *C*2/*m*. The large and small spheres represent La and B atoms, respectively.

LaB$_4$ adopts a tetragonal structure with *P*4/*mbm* symmetry under ambient conditions. A conventional cell of the *P*4/*mbm* phase contains four LaB$_4$ fu ($Z = 4$). La atoms occupy the Wyckoff 4*g* site, with three inequivalent boron atoms occupying the Wyckoff 4*e* site (B1), Wyckoff 4*h* site (B2), and Wyckoff 8*j* site (B3), respectively. The B1 and B3 atoms form an octahedron, as shown in Fig. 2(a). Four corners of each B-octahedron are connected by B2 atoms in the *a-b* plane, and the other two corners are connected with two identical neighbors through B1 atoms along the *c* axis. Thus, B atoms construct a three-dimensional network in which the La atoms are embedded. The lattice parameters calculated for the *P*4/*mbm* phase at 0 GPa ($a$ = 7.311 Å and $c$ = 4.185 Å) agree well with experimental data ($a$ = 7.324 Å and $c$ = 4.181 Å).[17] The B-octahedron is also found in similar compounds, such as YB$_4$, ThB$_4$, LuB$_4$, and UB$_4$.[42] It is noteworthy that these B-octahedrons have a similar volume, although the cell volume per fu of the compounds is different. For example, the volume per LaB$_4$ is 9.8% larger than that of YB$_4$, whereas the volume of the B-octahedron of LaB$_4$ is only 5.8% larger than that of YB$_4$.[43]

We further investigated the high-pressure behavior of the B-octahedron. At ambient pressure, the calculated B3-B3 and B3-B1 bond lengths are 1.838 Å and 1.790 Å, respectively, whereas the bond lengths of B1-B1, B2-B2, and B2-B3 are 1.724 Å, 1.830 Å, and 1.777 Å, respectively. Under the application of pressure, the bond lengths of B3-B3 and B3-B1 show a reduction of 8.28% and 8.70% at 100 GPa, whereas the B-B bond length outside the B-octahedron decreases by more than 10.46% at 100 GPa. Considering the fact that the B1-B1 bonds lie along the *c* axis and the B2-B2 and B2-B3 bonds are in the *a-b* plane, the lattice parameters *a* and *c* of *P*4/*mbm* exhibit different behaviors under pressure. The lattice parameter *a* decreases more quickly than the lattice parameter *c*; Fig. 3(b). Correspondingly, the volume of the B-octahedron decreases by 24.6% at 100 GPa, while the volume of the *P*4/*mbm* phase of LaB$_4$ decreases to 75.5% at 100 GPa, as shown in Fig. 3(a). Thus, the B-octahedron in the *P*4/*mbm* phase of LaB$_4$ is robust and stable.

Our calculations predict the *P*4/*mbm* phase transforms to an orthorhombic structure with *Pnma* symmetry (labeled the *Pnma*-I phase) at approximately 106 GPa. The *Pnma*-I phase of LaB$_4$ contains four fu in a conventional cell with four types of B atoms and one type of La atom. The B-octahedron that exists in the *P*4/*mbm* phase is destroyed and B atoms form six-membered rings in the *Pnma*-I phase. The six-membered B ring locates in two different planes, which cross each other. The six-membered B rings in the same plane share edges with two identical neighbors and the six-membered B rings in different planes are bonded with each other. Thus, the B atoms form a one-dimensional tube along the *b* axis, where La atoms are accommodated in the large voids of the tube structure. The optimized lattice parameters at 110 GPa are *a* = 7.321 Å, *b* = 2.771 Å, and *c* = 7.228 Å, respectively. The average inter- and intra-B-B bond lengths of the six-membered ring are 1.672 Å and 1.727 Å, respectively.

Interestingly, two metastable structures intensively compete for the low-enthalpy phase at 95 – 110 GPa. One of these structures also has orthorhombic *Pnma* symmetry (labeled *Pnma*-II), and the other has monoclinic *C*2/*m* symmetry. Indeed, the maximum difference in enthalpy between the *Pnma*-I, *Pnma*-II, and *C*2/*m* phases at 95 – 110 GPa is 0.240 eV/fu. These results indicate these three phases share some structural characteristics; specifically, the La atoms are accommodated in the large voids of a B framework. The conventional cell of the *Pnma*-II phase contains four LaB$_4$ units (*a* = 5.005 Å, *b* = 3.972 Å, *c* = 7.580 Å) at 110 GPa. Wyckoff 4*c* position is occupied by a La atom, with three inequivalent B1, B2, and B3 atoms occupying the Wyckoff 8*d*, 4*c*, and 4*c* positions, respectively. Five B atoms (B1 and B2) form a five-membered ring, which is located in two nearly perpendicular planes, as shown in Fig. 2(c). The five-membered B rings in the same plane share edges with each other and are connected to a B3 atom in a different plane, thus B atoms form a tubular structure that propagates along the *b* axis. The average bond length of the five-membered B ring at 110 GPa is 1.598 Å. The *C*2/*m* phase exhibits a *C*-centered monoclinic structure with six fu in each conventional cell (*Z* = 6). The B1 and B2 atoms occupy the Wyckoff $8j$ positions and the B3 and B4 atoms, the Wyckoff $4i$ positions. The B framework in the $C2/m$ phase is also a tube, which runs along the $b$ axis. The B1 and B2 atoms constitute a slightly buckled B plane with a bond distance of 0.41 Å. The six-membered B rings share edges with two identical six-membered rings and one square in the same plane. The average bond lengths of the six-membered B ring and square are 1.648 Å and 1.764 Å at 110 GPa, respectively. The buckled B plane is connected via B2 dumbbells, which consist of B3 and B4 atoms. The bond length of the B2 dumbbell is 1.660 Å at 110 GPa.

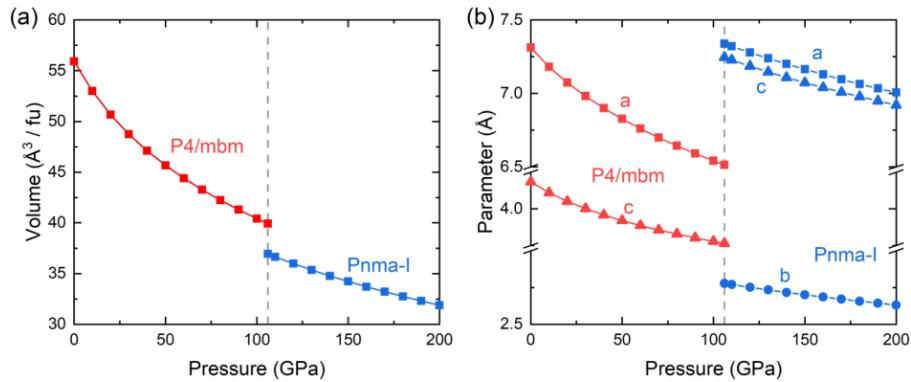

**Figure 3.** (a) Volume per formula unit versus pressure for LaB$_4$ and (b) lattice parameters for $P4/mbm$ and $Pnma$-I phase of LaB$_4$ as a function of pressure.

In order to further investigate these candidate structures, we assessed their mechanical stabilities. The calculated changes in the cell volumes and lattice constants over the studied pressure range are shown in Fig. 3. Obviously, the volume reduces significantly (by about 7.48%) when the phase transition occurs around 106 GPa (Fig. 3(a)). This large volume drop is due to a sudden change in atomic arrangement, i.e., the breakage of the B-octahedron. The discontinuous variation in volume as pressure increases across the phase boundary indicates a first-order phase transition. Intriguingly, the calculated volume of the $P4/mbm$ phase reduces as pressure increases, by 0.15 Å$^3$/GPa. However, an increase pressure reduces the volume of the $Pnma$-I phase more gradually, by only 0.05 Å$^3$/GPa. Overall, this data suggests that compression of LaB$_4$ becomes more difficult as pressure increases. Fig. 3(b) summarizes the lattice parameters for these competitive phases. For the tetragonal $P4/mbm$ phase, $a$ decreases with pressure by 0.008 Å/GPa, while $c$ decreases more gradually than $a$ (0.004 Å/GPa). The $a$ and $c$ lattice parameters for the $P4/mbm$ phase are 6.513 Å and 3.765 Å at 106 GPa, respectively. The lattice parameters of the $Pnma$-I phase can be split into two groups. The $a$ and $c$ lattice parameters are larger than $b$; $b$ is relatively more difficult to compress than *a* and *c*. The *a* and *c* decrease with pressure by 0.0035 Å/GPa, while pressure decreases *b* by 0.0016 Å/GPa. At 200 GPa, $a = 7.006$ Å, $b = 2.631$ Å, and $c = 6.921$ Å for the *Pnma*-I phase.

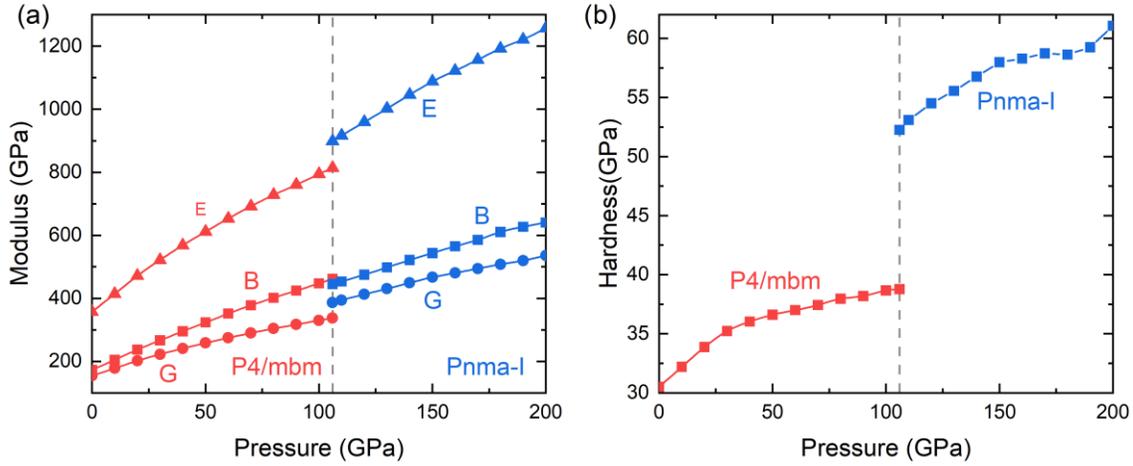

**Figure 4.** (a) Bulk modulus, shear modulus, and Young's modulus of $LaB_4$ and (b) hardness for the *P*4/*mbm* and *Pnma-I* phase of $LaB_4$ as a function of pressure.

Based on the Voigt-Reuss-Hill (VRH) approximation,[44-46] we estimated various mechanical moduli for $LaB_4$. The relationship between the mechanical moduli and pressure are shown in Fig. 4(a). The mechanical moduli of the *P*4/*mbm* and *Pnma*-I phases, such as bulk modulus (*B*), shear modulus (*E*), and Young's modulus (*G*), increase with pressure. Similar changes in the *E*, *B*, and *G* of the *P*4/*mbm* phase of $LaB_4$ are observed up to 106 GPa. Between 0 – 106 GPa, *E* increases from 357.3 to 814.0 GPa (127.8%), and *G* increases from 154.6 to 337.4 GPa (118.2%), though *B* significantly increases by 167.0% over the same pressure range. The *B*, *E*, and *G* values change suddenly when $LaB_4$ transforms from *P*4/*mbm* to *Pnma*-I (Fig. 4(a)). At 106 GPa, the *B* of the *Pnma*-I phase is 3.6% smaller than that of the *P*4/*mbm* phase, whereas *E* and *G* are 10.4% and 14.5% higher for *Pnma*-I than *P*4/*mbm*, respectively. The *E*, *B*, and *G* of the *Pnma*-I phase increase by 39.9%, 38.7%, and 40.0% over the pressure range from 106 – 200 GPa. At 200 GPa, *B*, *G*, and *E* are 640.78 GPa, 535.84 GPa, and 1257.11 GPa, respectively.

Vickers hardness is defined as the extent to which a solid material resists elastic or plastic deformation. The hardness value is determined experimentally by pressing an indenter into the surface of a material and measuring the size of the impression. There are many methods to predict the hardness of a material. We used the following formula to calculate the hardness of solid

materials:[47]

$$H_v = 2(k^2 G)^{0.585} - 3$$

where $k = G/B$ is the Pugh's modulus ratio. We found that LaB$_4$ has an unexpectedly high hardness value at 0 GPa. The *P*4/*mbm* phase of LaB$_4$ has a hardness of approximately 30.5 GPa, which is comparable to that of B$_4$C, SiO$_2$, ReB$_2$, and WC, and approximately one third of that of superhard diamond, indicating LaB$_4$ is a potential hard material. Below 50 GPa, the hardness of the *P*4/*mbm* phase increases quickly with pressure, as shown in Fig. 4(b). Between 0 to 50 GPa, the hardness of LaB$_4$ increases by 20.0%. The hardness increases more slowly above 50 GPa, as $G$ increases with pressure at a relatively low rate. At the phase transition pressure (106 GPa), the hardness of the *P*4/*mbm* and *Pnma*-I phases is 38.8 GPa and 61.1 GPa, respectively. This huge jump in the hardness of LaB$_4$ originates from the different $B$ and $G$ of these two phases. The hardness of the *Pnma*-I phase first increases quickly and reaches 58.0 GPa at 150 GPa, then plateaus up to 175 GPa, followed by another large increase up to 61.0 GPa at 200 GPa. This pressure-dependent hardness behavior comes from the interplay between $B$ and $G$ under pressure.

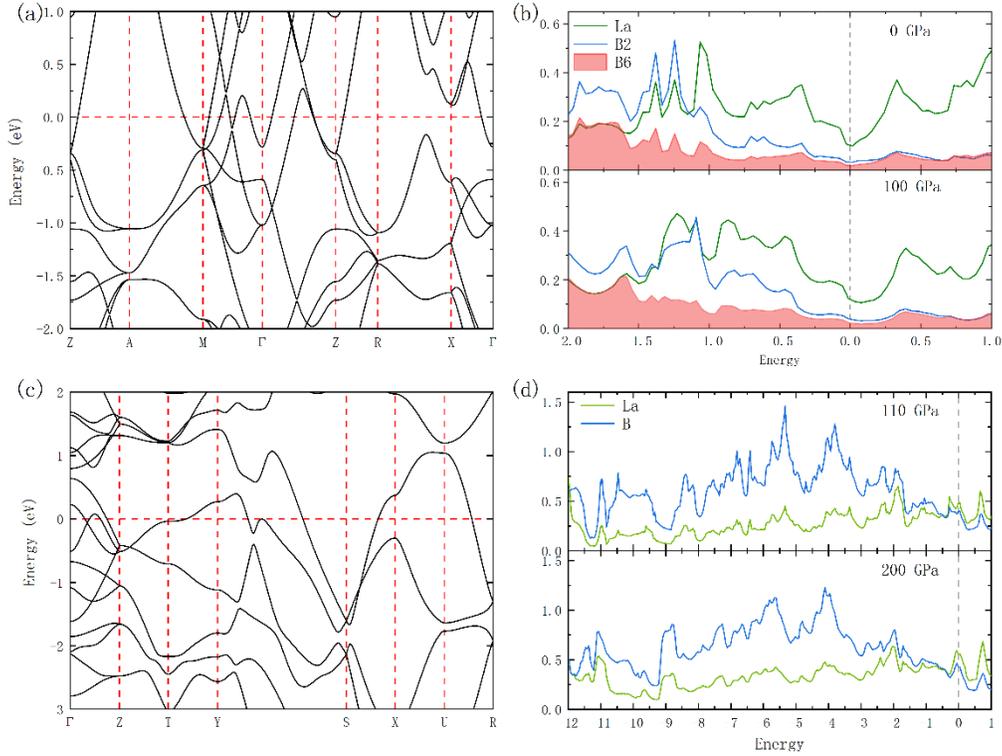

**Figure 5.** Electronic band structures of LaB$_4$ for (a) *P*4/*mbm* phase at 0 GPa and (c) *Pnma-I* phase

at 110 GPa. Electronic projected density of states (PDOS) of (b) *P*4/*mbm* phase at 0 GPa and 110 GPa and (d) *Pnma-I* phase at 110 GPa and 200 GPa. The unit of electronic density of states is states/eV/atom.

Electronic structure and chemical bonding determine the macroscopic properties of all materials, including their conductivity, elasticity, and hardness. We examined the effect of pressure on the electronic structure of LaB$_4$. Fig. 5 presents the electronic band structures for LaB$_4$ along the high-symmetry directions of the Brillouin zone and the corresponding density of states. The electronic band structures observed for *P*4/*mbm* and *Pnma*-I phases over their favored pressure ranges reveal LaB$_4$ exhibits metallic characteristics up to 200 GPa. The electronic projected density of states (PDOS) of the *P*4/*mbm* phase is shown in Fig. 5(b). Under ambient conditions, the La atom mainly contributes to the total DOS at the Fermi level. The B2 atom and B-octahedron contribute equally to the total DOS at the Fermi level. As pressure increases, hybridization between the PDOS of the B2 atom and B-octahedron increases [Fig. 5(b)], which accounts for the enhanced hardness of LaB$_4$. For the *Pnma*-I phase, three bands cross the Fermi level. Strong hybridization of La and B atoms occurs over a wide energy range; Fig. 5(d). La and B atoms contribute nearly equally to the DOS at the Fermi level at 110 GPa. The PDOS of the La atom at the Fermi level is 31.69% larger than that of the B atoms at the Fermi level at 200 GPa. The hybridization between the La and B atoms further enhances with increasing pressure.

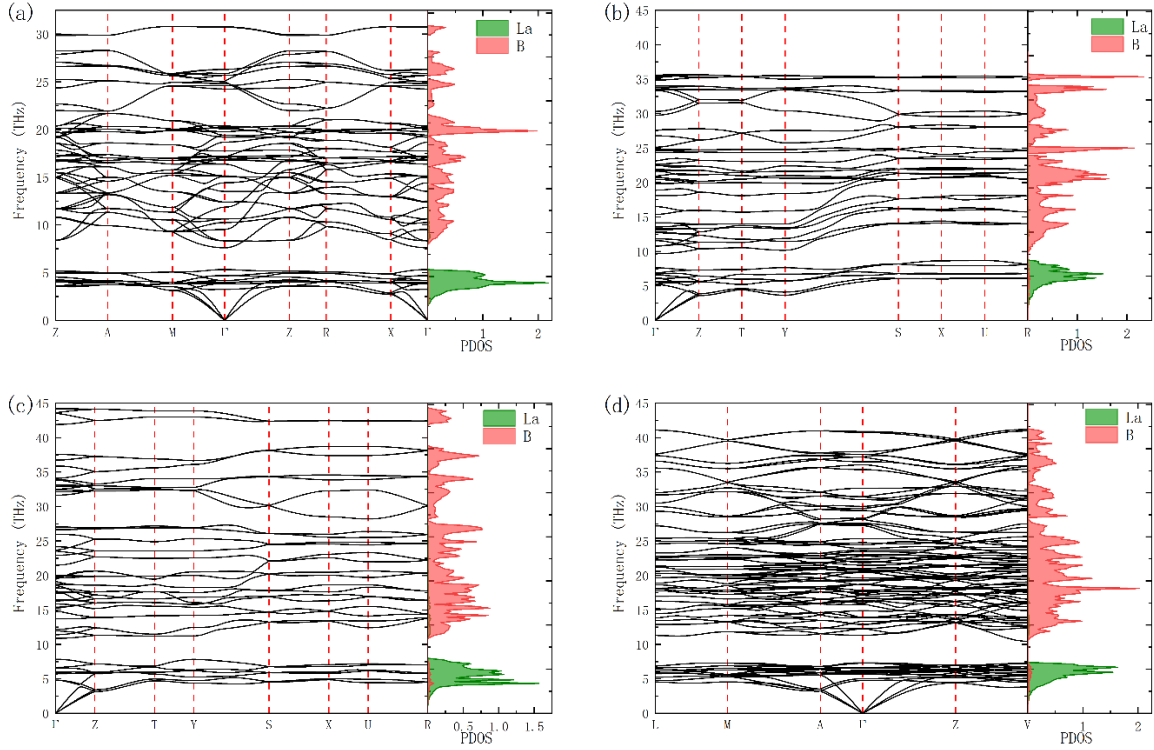

**Figure 6.** Phonon band structures and phonon projected density of states (PDOS) of LaB$_4$ for (a) *P*4/*mbm* at 0 GPa, (b) *Pnma*-I at 110 GPa, (c) *Pnma*-II at 110 GPa, and (d) *C*2/*m* at 110 GPa. The unit of phonon PDOS is states/THz/atom.

Fig. 6 presents the phonon band structures and projected DOS for *P*4/*mbm* at 0 GPa, *Pnma*-I at 110 GPa, *Pnma*-II at 110 GPa, and *C*2/*m* at 110 GPa. Additional phonon calculations confirmed *P*4/*mbm* and *Pnma*-I are thermodynamically stable over their favored pressure ranges, as indicated by an absence of imaginary frequency modes throughout the Brillouin zone. Two major phonon band structures were observed: vibration of the La atom dominates the low-frequency regions of all four phases, while vibrations of B atoms dominate the high-frequency region, as shown in Fig. 6. The relatively low-frequency modes (below 10 THz) in the *P*4/*mbm* phase at ambient conditions are mainly due the La atom, with a sharp peak near 4 THz; the higher frequency range (over 10 THz) is dominated by B atoms, with a sharp peak near 20 THz. The vibration modes of B and La atoms increase with pressure. The vibration modes of the B atoms disperse up to 35 THz in *Pnma*-I at 110 GPa, and those of the La atom disperse up to 7.5 THz, as shown Fig. 6 (b). The maxima of the high-frequency regions of the two metastable phases *Pnma*-II and *C*2/*m* are 44.35 THz and 41.27 THz, respectively, larger than that of the *Pnma*-I phase; this result can be attributed to the different atomic arrangements.

## 4. Conclusions

We systematically explored the pressure-dependent phase transitions and mechanical properties of LaB$_4$ via first-principles calculations. When compressed, the *P*4/*mbm* phase transforms to the *Pnma*-I phase at 106 GPa, and this phase transition is of the first order. Two metastable phases (*Pnam*-II and *C*2/*m*) share some structural features with the stable *Pnma*-I phase at high pressure. The electronic band structures suggest these competitive phases (two stable and two metastable) are metallic. Furthermore, the *P*4/*mbm* phase of LaB$_4$ possesses a high hardness of 30.5 GPa under ambient conditions, which originates from La and B hybridization. Phonon calculations indicate the *P*4/*mbm* and *Pnma*-I phases are thermodynamically stable, as there are no imaginary frequencies within their favored pressure ranges. Overall, this study demonstrates LaB$_4$ is a hard material and that its high-pressure behaviors may hold significant potential for the application of rare-earth metal borides.

## Acknowledgements

This work was supported by the National Natural Science Foundation of China (Grants Nos. 11874318 and 11774299), the Natural Science Foundation of Shandong Province (Grants Nos. ZR2018MA043 and ZR2017MA033).